# A Fluorescent Ratiometric Potassium Sensor Based on IPG4-Silica Microparticles for Selective Detection and Fluorescence Imaging of Potassium Cations

Francesco Colella,[a,b] Stefania Forciniti,[a] Valentina Onesto,[a] Giuliana Grasso,[a] Helena Iuele,[a] Giuseppe Gigli[a,c] and Loretta L. del Mercato[*,a]

Potassium cations play many important roles in living organisms, especially in electro-physiology, since they are involved in neurotransmission and muscle contractions. We report the synthesis of a ratiometric fluorescent microsensor for potassium ($K^+$) detection, based on the fluorescent probe ION potassium green 4. Potassium-sensitive fluorescent microparticles were obtained by using silica as core material. We obtained silica-based microsensors with a size in the micrometer range, spherical shape, good monodispersity, optimal selectivity and a sensitivity range between 0 to 40 mM. The microsensors also proved to be non-toxic in cell cultures as well as suitable for fluorescent imaging, offering new possibilities for non-invasive optical detection, quantification and in situ monitoring of $K^+$ variations in cell culture systems.

## Introduction

Potassium cations ($K^+$) are among the most important inorganic ions for electro-physiological processes, being essential for both muscles contraction and neurons communication [1]. To maintain cell membrane potentials, the $K^+$ ions concentration in plasma and interstitial fluids is almost 100 times smaller compared to the intracellular one (extracellular 3-5 mM, intracellular 100-150 mM) [1]. Such concentration difference is essential for animal physiology and is kept constant by the activity of $Na^+/K^+$ ATPase pump and kidney filtration [1]. Dysregulations in $K^+$ homeostasis are involved in pathophysiological events, mainly cardiac (i.e. cardiac arrhythmia) and central nervous system dysfunctions (i.e. migraine and epilepsy) [1–3]. There is also evidence that $K^+$ ions accumulation plays a relevant role in cancer, by creating a hostile environment for immune cells, preventing them from attacking the tumor [4,5]. Considered that dysregulations in $K^+$ concentration are involved in different pathological phenomena, the development of new tools for precisely studying and detecting their variations in *in vitro* and *in vivo* models is still a hot topic. Among the most common techniques for measuring cations variations in biological systems, ion-selective electrodes (ISEs) and microelectrode arrays (MEAs) are largely used for sensing ion concentrations, including $K^+$. Nonetheless, they have limits related to cation selectivity [6], low sensitivity, potential interference from other ions, circumscribed spatial resolution (e.g., only $K^+$ concentrations in the close proximity to the electrode can be detected), limited measurement duration and complexity of data interpretation [7]. An alternative approach to the electrochemical sensors is the ion-selective optodes (ISOs), which are the optical equivalent of an electrode, usually based on optical fibers [8,9]. Meanwhile, patch-clamp is largely used to measure ions fluxes through cell membranes in biological samples [10,11]. Optical methods for cations analyses are based on fluorescent organic probes, which are able to bind cations and give a quantitative response resulting from a fluorescence enhancement or quenching of the emission intensity [12–16].

However, the selectivity between different cations is an issue even in this case [14]. New molecular approaches (i.e., probes and fast-responding chemo- and nanosensors) are always under development. These fluorescent probes can be ratiometric or non-ratiometric. In a ratiometric sensor the final readout is given by the ratio between the sensitive signal and a non-sensitive signal. Furthermore, a molecular probe can be intrinsically ratiometric, with two fluorescence emission signals, and one of them insensitive to the analyte. In the case of non-ratiometric probes, a second fluorescent dye can be used as reference signal [15–18]. In the last years several prototypes of $K^+$ sensing micro/nano-particles were reported. One of the approaches more similar to the one we adopted in this study was reported by MacGilvary *et al.*, which utilized bovin serum albumin (BSA) coated silica microparticles to covalently bind IPG4 (ION Potassium Green 4, ION Biosciences, United States) as sensitive dye and Alexa Fluor 594 as reference dye. The result was a ratiometric microsensor with a sensitivity for $K^+$ concentration from 0 to 128 mM.[19] Xie, Xiaojiang *et al.* reported the synthesis of fluorescent $K^+$ sensitive polystyrene microparticles. However, despite these microsensors show good selectivity towards $K^+$ over $Na^+$, $Li^+$ and bivalent cations ($Ca^{2+}$ and $Mg^{2+}$), the sensitivity range of the microsensors is from 0 to 60 µM, which is too low for monitoring [$K^+$] in biological samples [20]. Lee Chang H. *et al.* reported the synthesis of dual-mode $K^+$ sensing nanoparticles that can be used as fluorescent ratiometric sensors or for photoacoustic imaging. These nanosensors displayed sensitivity from 1 mM to 1 M, that could be suitable for monitoring both intracellular and extracellular $K^+$ variations. However, the system show interference with other cations, especially $Na^+$ [21]. One of the most commonly used molecular probes for measuring $K^+$ concentration changes in living cells is the potassium-binding benzofuran isophtalate (PBFI) [22–25], which has major limitations. PBFI presents limited specificity: it is only 1.5-fold more selective toward $K^+$ compared to sodium cations ($Na^+$), which makes it more suitable for intracellular sensing (where $Na^+$ concentration is only 5-30 mM), it displays weak fluorescence and it is difficult to load intracellularly [26]. The other limit is that PBFI needs to be excited in the far UV region, with a laser source of 350 nm, which might have harmful effects on living cells [27].

In this study, we produced a stable and highly sensitive optical sensor for K+ detection. To overcome the limitations of the current sensors, we chose the Asante potassium green 4 (APG-4), lately renamed ION Potassium Green 4 (IPG4) (ION Biosciences, United States), which is reported to be the most selective, currently available, fluorescent probe for K+, since it is declared to be 100 times more selective toward K+ compared to Na+. The manufacturer declares a sensitivity range between 0 – 38 mM with a saturation kinetic and a $K_d$ of 7 mM [28]. The IPG4 fluorescence emission displays an increase correlated to the increase on K+ concentration. The fluorescence emission is in the green region of the visible light, and the declared $\lambda_{ex}$ max is at 525 nm, although it can be also excited with a 488 nm laser line. After, we characterized the physicochemical properties of the sensor and we assessed its biocompatibility and application in a cell culture system, we are proposing a fluorescent ratiometric microsensor for applications in high resolution imaging of biological models, allowing a real-time monitoring of the K+ variations.

## Experimental section

### Synthesis of the microsensors

A solution composed by 60 mL of EtOH, 7 mL of MilliQ water, 9 mL of NH4OH 28%, 15 mg of KCl was poured into a 250 mL reaction flask, sealed with a rubber septum and kept under stirring at 250 rpm. A solution composed of 35 mL of anhydrous ethanol and 2.38 mL of TEOS was injected into the reaction flask by using a syringe pump, with a flowrate of 0.12 mL/min. After the injection the reaction was kept under stirring for 12 hours at 250 rpm at room temperature.

The obtained silica microparticles were purified by means of centrifugation at 2500 rpm for 5 minutes at 21 °C. The pellet was suspended in 50 mL ethanol and centrifuged again at 2500 rpm for 5 minutes at 21°C. The washing steps were repeated 3 times. For the synthesis of aminated microparticles, 500 mg of the obtained silica microparticles core were suspended in 30 mL of anhydrous ethanol and sonicated in a sonication bath for 10 minutes. The microparticles suspension was poured in a 100 mL reaction flask and sealed with a rubber septum. A solution composed of 10 mL of anhydrous ethanol, 250 µL of TEOS and 50 µL of APTES was injected in the reaction flask through a syringe pump, at the flowrate of 0.05 mL/min. The reaction was kept under stirring at 300 rpm, at RT, for 12 hours. The aminated microparticles were then washed by centrifugation 3 times (2500rpm, 5 minutes, 21°C). The ninhydrin assay was used to check the successful aminosilanization on the plain microparticle surface (see ninhydrin assay).

The SiO2@IPG4 MPs were obtained through the amide coupling performed with HATU. 100 mg of SiO2@NH2 were suspended in DMF and centrifuged 3 times at 8000 rpm for 4 minutes on a benchtop centrifuge, in order to remove the residual ethanol and water. The microparticles were then suspended in 1 mL of DMF. IPG4 lyophilized powder was solubilized in DMF and the corresponding volume for 50 µg of IPG4 was diluted in 100 µL of DMF. HATU was solubilized in DMF at known concentration and the corresponding volume for 1.9 molar equivalent (26.6 µg) was added to the IPG4 solution. 10 µL of DIPEA were added as base catalyst to the solution, and the solution was kept still in the darkness for 30 minutes (in order to obtain a complete activation of the carboxylic functions). The activated IPG4 solution was added to the silica MPs suspension, DMF was added to a final reaction volume of 2 mL and the reaction was kept under stirring for 1 hour. The resulting microparticles were then washed by centrifugation in ethanol 3 times (2500rpm, 5 minutes, 21°C).

The SiO2@RBITC-IPG4 MPs were obtained by forming an outer shell of RBITC around silica, with a slightly different protocol. 6 mg of RBITC were dissolved in 2 mL of DMF and 50 µL of APTES were added to the solution. The reaction was kept under stirring for 3 hours (600 rpm, RT). 500 mg of silica microparticles core were suspended in 30 mL of ethanol and sonicated in a sonication bath for 10 minutes. The microparticles suspension was poured in a 100 mL reaction flask and sealed with a rubber septum. 8 mL of anhydrous ethanol, 250 µL of TEOS and 50 µL of APTES were added to the previous solution containing RBITC-APTES. The solution was injected into the reaction flask through a syringe pump, at the flowrate of 0.05 mL/min. The reaction was kept under stirring at 300 rpm, at RT, for 12 hours. The microparticles were then washed by centrifugation for 3 times (2500rpm, 5 minutes, 21°C). The ninhydrin assay was used to check if the microparticles were aminated (see Ninhydrin assay). The last step for the conjugation of the IPG4 to obtain the SiO2@RBITC-IPG4 is the same as previously described for the synthesis of SiO2@IPG4.

The same protocol for the synthesis of SiO2@RBITC-IPG4 was also used for the synthesis of SiO2@Cy3-IPG4, by switching RBITC with Cy3.

### Ninhydrin assay

2 mg of dry silica microparticles were suspended in 1 mL of ethanol in a 1.5 mL plastic tube. 100 µL of ninhydrin solution (2% w/v in ethanol) were added to the solution. Plastic tubes were placed in a thermoshaker and let react for 20 minutes at 90°C, 200 rpm. The tubes were cooled down in an ice-cold bath and then centrifuged at 8000 rpm for 5 minutes. The supernatant was collected and the absorbance at 570 nm was measured with CLARIOstar Plus plate reader (BMG LABTECH, Ortenberg, Germany). A solution without microparticles was used as negative control and five glycine solutions at known concentration were used to plot the calibration curve.

### Physical characterization

Size and morphology of the microsensors were characterized by means of scanning electron microscopy (SEM, Sigma 300VP, Zeiss, Germany) (accelerating voltage of 5 kV, secondary electron detector (SE2), 5000x, 10000x and 30000x magnifications). All the samples were sputter-coated (compact coating unit CCU-010, SafeMatic GmbH, Zizers, Switzerland) with a 10 nm thick gold layer (Target Au Ø 54 mm×0.2 mm; purity, 99.99%) prior to their observation under the microscope. The mean diameter of the microparticles was obtained with Fiji Imagej (https://imagej.net/software/fiji/.net, NIH, USA), based on the SEM diameter of 100 microparticles.

The hydrodynamic diameter and the surface charge of the microparticles was assessed by means of dynamic light scattering (DLS) and zeta potential (ZP) analysis using Zetasizer Nano ZS90 (Malvern Instruments, Malvern, Worcestershire, UK) equipped with a 4.0-mW He-Ne laser operating at 633 nm, while an avalanche photodiode detector was used. Semi-micro cuvette (BRAND®, PMMA, minimum filling volume 1.5 mL, #BR759115) was used for DLS analysis. Disposable folded capillary ζ Cell (Malvern Instruments, Malvern, Worcestershire, UK; #DTS1070) was used for loading the sample. Deionized water was used as the dispersant (n = 1.33, η = 0.88), and measurements were performed at 25 °C. The refractive index used during the acquisitions was taken as n=1.54 and absorption as k=0.00. The microsensors stock solutions (40 mg/mL) were diluted in ultrapure water, by pouring 10 µL of stock solution in 1 mL of MilliQ water. The suspensions were mixed with a vortex and then sonicated in a sonication bath for 10 minutes at 40kHz.

**Calibration and stability evaluation of the microsensors**

The fluorimetric calibrations of the IPG4 dye, the IPG4 conjugated silica microparticles and the different prototypes of the ratiometric microsensors were all performed in [$K^+$] adjusted MES buffer (50mM, pH 7.5) in spectral scan mode by using CLARIOstar Plus plate reader (BMG LABTECH, Ortenberg, Germany). All experiments were performed in black 96 well plates (Corning® 96-well Black Flat Bottom Polystyrene NBS Microplate, Corning, Glendale, Arizona, USA; #3650). To evaluate the stability of the sensing probes, the ratiometric fluorescent microparticles (2 µL, stock solution 40 mg/mL) were incubated with different concentrations of $K^+$ (0, 10 mM, 20 mM, and 40 mM) in MES buffer. The microplate was stored in dark, at 4°C for 7 days. The fluorescence read out was monitored and collected at day 1, 3, 5 and 7. The $K^+$ detection was obtained via ratiometric calculation obtained by the emission peak integration. The fluorescence spectra of the microsensors were extracted as follows: for IPG4, $\lambda_{ex}$= 488 nm and $\lambda_{em}$= 510-570 nm; for RBITC, $\lambda_{ex}$= 555 nm and $\lambda_{em}$= 570-700 nm.

The CLSM-based calibration of the microsensors was performed with a ZEISS LSM-700 (Carl Zeiss, Jena, Germany) by dispersing the $K^+$ sensitive microparticles in [$K^+$] adjusted MES buffer (50mM, pH 7.5). Samples were immobilized in a µ-Slide 8 Well Ibidi® chamber plate (Ibidi GmbH, Gräfelfing, Germany, #80827), previously functionalized with poly-lysine (0.3% in water). Micrographs were acquired using a PlanApochromatic 63X/1.4 oil DIC objective (213.39 µm x 213.39 µm) and a resolution of 1024x1024 pixel. The 488 nm Argon laser line and the 555 nm laser line were used as excitation sources.

**Image analysis**

CLSM images were analysed using a previously developed algorithm [12,13] for the automatic seg-mentation and extraction of the fluorescence intensities experienced by the single microparticles. Briefly, reference channel images were pre-processed and binarized removing noise in the images, small holes in the objects, and suppressing structures that were connected to the image border. Then images were segmented by a watershed transformation [29] identifying single particles in the binarized reference channel. This image was used as a mask to extract, for each microparticle, the pixel-by-pixel ratio of the fluorescence intensities between indicator and reference channels. Finally, mean and standard deviation among all the considered sensors were determined.

**Biocompatibility evaluation**

The cytotoxicity of $SiO_2$@RBITC-IPG4 microparticles was evaluated through CellTiter-Glo® Cell Viability Assay (Promega Corporation, Germany) on breast cancer cells MCF-7 (HTB-22™; ATCC, Rockville, USA), human pancreatic cancer cells PANC-1 (CRL-1469™; ATCC, Rockville, USA), melanoma cell line SKMEL-2 (HTB-68™; ATCC, Rockville, MD, USA), and 3T3 murine fibroblasts (CCL-163™ ; ATCC, Rockville, USA) cultured at 37 °C in a humidified 5% $CO_2$ incubator. SK-MEL-2 were cultured in Eagle's Minimum Essential Medium (MEM, Sigma-Merck KGaA, Darmstadt, Germany) whereas Panc-1, and MCF-7 and 3T3 fibroblasts were grown in Dulbecco's Modified Eagle's Medium (DMEM,) Sigma-Merck KGaA, Darmstadt, Germany), supplemented with 10% Fetal Bovine Serum (FBS, Gibco, Thermo Fisher Scientific Inc., Waltham, MA, USA), 2 mM L-glutamine, and 100 U/mL penicillin and streptomycin (Sigma-Merck KGaA, Darmstadt, Germany). $5x10^3$ cells/well were seeded in 96-well plates and treated with or without $K^+$ sensing microparticles at the final concentrations of 0.05, 0.1 and 0.3 mg/mL. Cell viability based on the quantification of ATP produced by metabolically active cells, was measured at 0, 24 and 48 hours by adding a volume of CellTiter-Glo® reagent equal to the volume of medium present in each well. Then, the samples were mixed for 5 minutes to induce cell lysis and to allow the extraction of ATP from live cells. After, samples were incubated for additional 25 minutes at room temperature to equilibrate the luminescent signal. Luminescence was recorded using CLARIOstar Plus plate reader (BMG LABTECH, Ortenberg, Germany). The viability of untreated cells was used as control and the luminescence values of medium containing sensors were subtracted from those of the treated or untreated samples.

**Evaluation of microsensor properties in cell culture**

SK-MEL2 tumor cells were counted by trypan blue dye exclusion and $6x10^4$ cells/well were seeded into a µ-Slide 4 Well (Ibidi® GmbH, Gräfelfing, Germany). After 24 hours, cells were stained with a nuclear marker, Hoechst 33342 (B2261, Sigma Aldrich) for 20 minutes, washed in PBS 1X and treated with $K^+$ sensing microparticles at the final concentration of 0.05 mg/ml. Then, cells were stimulated with 1 µM of nigericin for 1 hour at 37°C in order to induce $K^+$ efflux by forming pores on the plasma membrane. Untreated cells were used as a control. Representative images were acquired using a CLSM (LSM 980, ZEISS, Germany) equipped with a 40x oil-immersion objective. The maximum projections of z-stack images were obtained using Image J software. For the fluorometric measurements of K$^+$ concentration in the cell culture media, SKMEL-2 cells (5x10$^3$ cells/well) were seeded in 96-well plates and treated with or without 1 µM of nigericin for 1 hour. Then, the cell culture supernatant was collected for each conditions and added with K$^+$ sensing microparticles at the final concentration of 0.5 mg/ml. Samples were analyzed in a black 96-multiwell plate through a CLARIOstar plate reader (BMG Labtech, Ortenberg, Germany) by using a spectral scan mode.

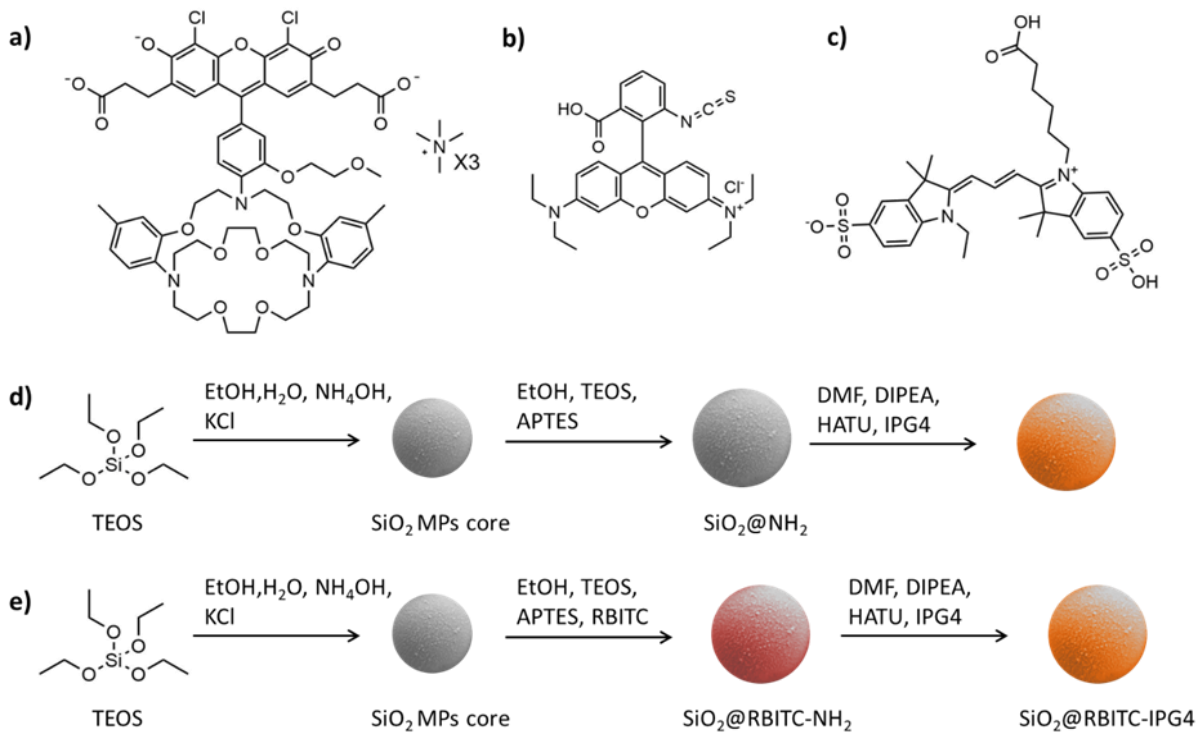

**Figure 1.** Molecular structure of the employed dye and synthetic scheme for the synthesis of the silica microsensors: **a)** molecular structure of IPG4 TMA salt; **b)** molecular structure of the rhodamine B isothiocyanate; **c)** molecular structure of cyanine 3; **d)** synthesis scheme with reaction conditions of the SiO$_2$@IPG4; **e)** synthesis scheme with reaction condition of the SiO$_2$@RBITC-IPG4.

**Statistical analysis**

The experiments were performed in triplicate, and the results were reported as the mean ± standard error unless otherwise stated. The limit of detection (LOD) was calculated as LOD = 3σ/s, where σ is the standard deviation of blank measurements (without K$^+$) and s is the slope of ratio versus K$^+$ concentration. Statistical differences were considered significant at p<0.05 using two-way analysis of variance.

## Results

**Design and synthesis of the fluorescent ratiometric microsensors**

In order to design a fluorescent ratiometric K$^+$ microsensor two fluorescent signals are required, one that must be analyte-sensitive and the second that must be non-sensitive to the analyte. Some organic fluorescent probes are intrinsically ratiometric (e.g., pyranine as pH-sensitive probes [17,30]), but IPG4 is a non-ratiometric probe for K$^+$ detection. Therefore, we have to supply the microparticles with a second fluorescent signal, that will provide an inner reference signal. We selected different fluorescent dyes to be used in combination with IPG4. The main requirement for the reference dye was the fluorescence emission in a different region of the visible spectra, avoiding any overlap with the emission peak of IPG4. Based on the known spectral properties, we used different fluorophores as potential candidates (as listed in Table S1). We synthesized silica particles as starting material for the proposed ratiometric microsensors, due to its inertness and the customizable chemical properties [31]. Silica microparticles were produced by using a modified Stöber reaction, by exploiting TEOS hydrolysis in alkaline hydroalcoholic solution [32,33]. This method allowed the reproducible and scalable production of silica microparticles with fine control over size and shape by tuning the reaction conditions [17,32]. The obtained silica microparticles (with a size of approximately 1.5+0.4 µm, see Figure 3 and Table 1) were silanized by using APTES in anhydrous ethanol. The presence of free primary amines on the surface of the microparticles (due to the APTES silanization) was confirmed and quantified by using the ninhydrin colorimetric assay (see Table S2 and Figure S1).

The next step was the conjugation of IPG4 on aminated silica microparticles. IPG4 TMA salt was used for the functionalization, by exploiting the two carboxylic groups, which characterize the molecular structure of the probe, to obtain an amide bond on the microparticles surface (see Figure 1a). All the attempt of using

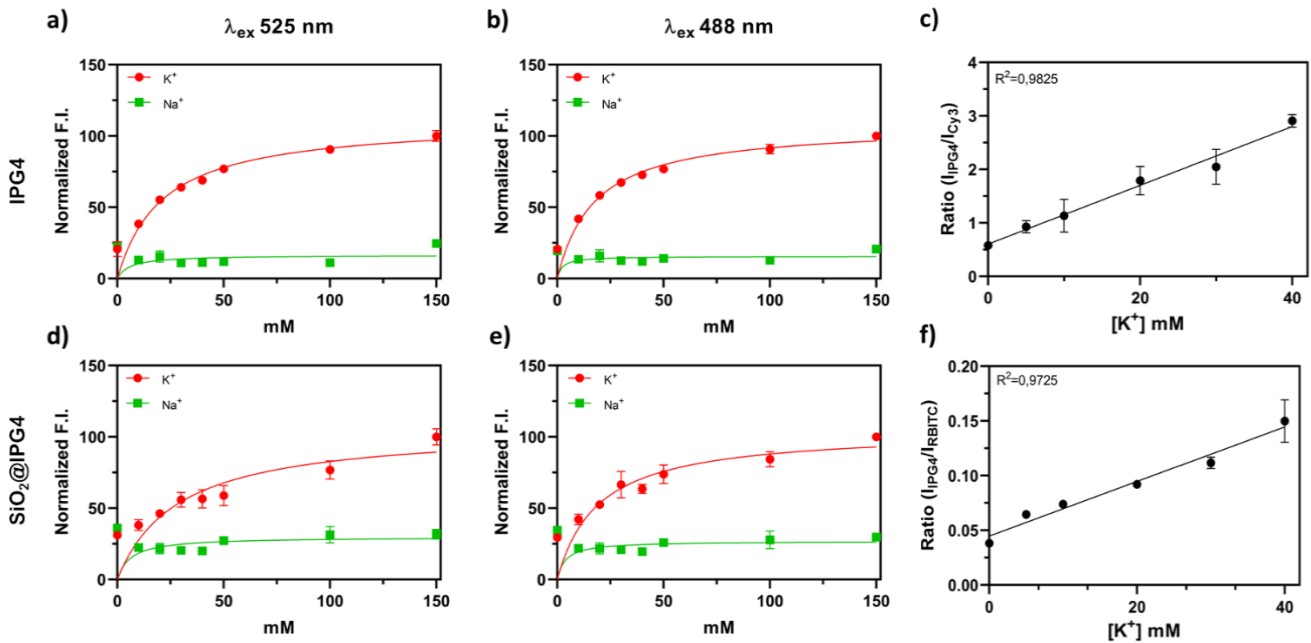

**Figure 2.** Kinetic analyses of IPG4 in presence of K$^+$ and Na$^+$ and ratiometric calibration curves of the developed microsensors. **a)** correlation between K$^+$ and Na$^+$ concentration and the fluorescence intensity of IPG4 dissolved in MES buffer, in the range between 0-150 mM ($\lambda_{ex}$ 488 nm and $\lambda_{ex}$ 525 nm, $\lambda_{em}$ 545-650 nm); **b)** correlation between K$^+$ and Na$^+$ concentration and the fluorescence intensity of IPG4 in the range between 0-150 mM ($\lambda_{ex}$ 488 nm and $\lambda_{ex}$ 525 nm, $\lambda_{em}$ 545-650 nm); **c)** ratiometric calibration of SiO$_2$@Cy3-IPG4 in the range of [K$^+$] = 0-40 mM ($\lambda_{ex}$ 488 nm, $\lambda_{em}$ 545-650 nm); **d)** correlation between K$^+$ and Na$^+$ concentration and the fluorescence intensity of SiO$_2$@IPG4 in the range between 0-150 mM ($\lambda_{ex}$ 488 nm and $\lambda_{ex}$ 525 nm, $\lambda_{em}$ 545-650 nm); **e)** correlation between K$^+$ and Na$^+$ concentration and the fluorescence intensity of SiO$_2$@IPG4 in the range between 0-150 mM ($\lambda_{ex}$ 488 nm and $\lambda_{ex}$ 525 nm, $\lambda_{em}$ 545-650 nm); **f)** ratiometric calibration of SiO$_2$@RBITC-IPG4 in the range of [K$^+$] = 0-40 mM, with excitation ($\lambda_{ex}$ 488 nm, $\lambda_{em}$ 545-650 nm). All the samples were dissolved or suspended in MES buffer (50 mM, pH 7.5). The analysis was performed on a CLARIOstar plate reader by using spectral scan mode, in a black 96 multi well plate.

EDC as coupling agent produced no results. So, we performed the amide synthesis by using HATU, as coupling agent, along with DIPEA in DMF as solvent (see Figure 1d). Following the IPG4 immobilization step, the microparticles displayed a bright pink color, indicating the binding of the probe on SiO$_2$ microparticles. The correct conjugation of IPG4 on the silica surface was further confirmed by fluorimetric calibration and kinetic calibration (see Optical and kinetic characterization).

The next step in the synthesis of the ratiometric microsensors is the immobilization of two fluorophores on the silica surface. The strategy that we adopted is based on the formation of an outer shell of the reference dye on the surface of microparticle, with an excess of APTES that should supply the free primary amino groups for the further conjugation of IPG4 (see Figure 1e). All the selected dyes, listed in Table S1, were used in this way to produce a ratiometric system. Only two of the selected reference dyes showed the correct properties for the integration in the ratiometric scheme, which were the RBITC and the Cy3 (the molecular structures of the molecules are shown in Figure 1b and 1c). However, all these molecules showed FRET phenomenon with IPG4, when immobilized on the surface of silica microparticles (see Table S1).

**Table 1. DLS analyses of the obtained microsensors**

|   | Diameter (μm) | PDI | ζ-potential (mV) |
|---|---|---|---|
| SiO$_2$@IPG4 | 1.66±0.11 | 0.333 | -9.41±0.97 |
| SiO$_2$@Cy3-IPG4 | 1.57±0.05 | 0.230 | -1.26±0.67 |
| SiO$_2$@RBITC-IPG4 | 1.96±0.16 | 0.338 | -10.2±0.68 |

## Optical and kinetic characterization

In order to evaluate the successful binding of IPG4 on the surface of the silica microparticles and their selectivity towards K$^+$ ions, the optical properties of the IPG4 probe itself and IPG4-conjugated silica microbeads were analyzed. The emission spectra of IPG4 with different concentrations of K$^+$ (by using KCl dissolved in MES buffer) and with different concentrations of Na$^+$ (by using NaCl dissolved in MES buffer) were recorded with two excitation wavelengths (488 nm and 525 nm). Based on the physiological range of concentration of Na$^+$ and K$^+$ in the interstitial fluids, relatively 100-150 mM for Na$^+$ and 3-5 mM for K$^+$, the sensitivity of IPG4 in the cations range of concentration from 0 to 150 mM was explored.

As shown in Figure 2a and 2b, the IPG4 dye displayed sensitivity only for the K$^+$ with no sensitivity toward Na$^+$ in the 0-150 mM range. Notably, the same sensitivity was retained following the immobilization of the fluorescent probe on the silica microparticles (see Figure 2d and 2e), likely showing that the immobilization steps did not affect the properties of the IPG4 probe.

The dissociation constant ($K_d$) and maximum specific binding ($B_{max}$) for K$^+$ and Na$^+$ were calculated for both IPG4 and IPG4 conjugated on silica microparticles. As reported in Table S3, the calculated $K_d$ for K$^+$ may vary based on the excitation wavelength and sensing platform (declared K$_d$= 7 mM, ION Biosciences), while the B$_{max}$ was comparable in all the systems (~105 mM). All the sensing systems, with both excitation wavelength, showed low binding capacity for sodium cations (Na$^+$) (Figure 2), with a B$_{max}$ approximately around 20 mM. Furthermore, there were no major differences in the results obtained with λ$_{ex}$ 488 nm and λ$_{ex}$ 525 nm (Figure 2 and Table S3), proving that the IPG4 probe was suitable for the K$^+$ mapping by using the 488 nm Argon laser line exploited in confocal laser scanning microscopy (CLSM).

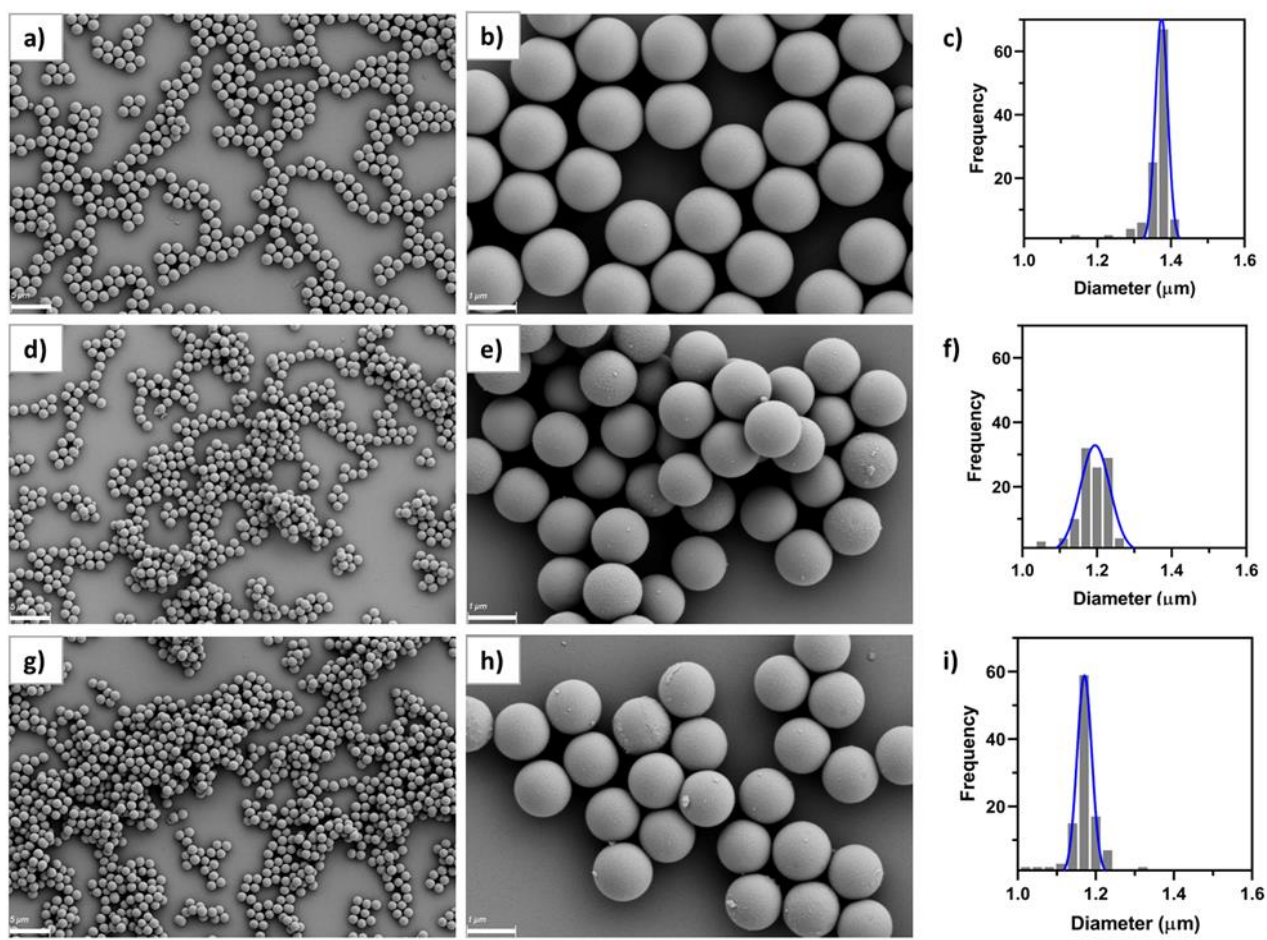

**Figure 3. Morphological characterization of K$^+$-sensors.** SEM micrographs and size distribution of **a)** SiO$_2$@IPG4 acquired with 5000X magnification; **b)** SiO$_2$@IPG4 acquired with 30000X magnification; **c)** SiO$_2$@IPG4 size distribution chart based on SEM micrograph, mean diameter =1.374 ± 0.016 µm; **d)** SiO$_2$@Cy3-IPG4 acquired with a 5000X magnification; **e)** SiO$_2$@Cy3-IPG4 acquired with 30000X magnification; **f)** SiO$_2$@Cy3-IPG4 size distribution chart based on SEM micrograph, mean diameter =1.195 ± 0.039 µm; **g)** SiO$_2$@RBITC-IPG4 acquired with 5000X magnification; **h)** SiO$_2$@RBITC-IPG4 acquired with 30000X magnification; **i)** SiO$_2$@RBITC-IPG4 size distribution chart based on SEM micrograph, mean diameter =1.171 ± 0.018 µm.

The calibration curves were performed for the different prototypes of ratiometric microsensors. However, only two prototypes showed the proper optical properties for the ratiometric system, namely the SiO$_2$@RBITC-IPG4 and the SiO$_2$@Cy3-IPG4. All the prototypes showed FRET phenomenon between the dyes, due to the close proximity on the silica surface. Among the selected dyes, the 7ACC1 had the maximum fluorescence emission before the λ$_{ex}$ max of IPG4. However, 7ACC1 displayed a different λ$_{em}$ max when covalently bound on silica, with an emission shift to lower wavelength, between 500-600 nm (see Figure S2), which overlapped with the fluorescence emission of IPG4. We hypothesized that spectra red-shift could be related to the conversion of the free carboxylic function to amide. Hence, the microparticles defined as SiO$_2$@IPG4-7ACC1 resulted to be not suitable as ratiometric K$^+$ microsensor.

The emission spectra of SiO$_2$@A594-IPG4 also showed obvious FRET [34] phenomenon. Indeed, the emission peak of A594 was the only one detectable after the conjugation of both fluorophores on silica microparticles, at three different excitation wavelengths (488 nm, 525 nm and 555 nm). Unexpectedly, SiO$_2$@A594-IPG4 showed an inverted correlation towards increasing [K$^+$] and the fluorescence intensity. The enhanced fluorescence of IPG4 related to K$^+$ seemed to act as quencher on A594 fluorescence intensity, with a sensitivity range between 0 and 40 mM, perfectly comparable with the properties of IPG4 (see Figure S3). However, despite the microparticles showed sensitivity to K$^+$, they could not be used as ratiometric optical sensors due to the absence of a reference signal.

Only the microparticles carrying RBITC and Cy3 as reference dyes resulted to be suitable as ratiometric K$^+$ microsensors by means of spectrofluorimetric analyses, as shown in Figure 2c and 2f. These two ratiometric microsensors showed sensitivity in the range between 0-40 mM of K$^+$, in accordance to the results obtained with the free IPG4 probe.

in the range from 0 to 40 mM and the results were reported in Figure 4. As shown in Figure 4a and 4b, the SiO$_2$@RBITC-IPG4 microsensors displayed a linear response towards K$^+$ in the range

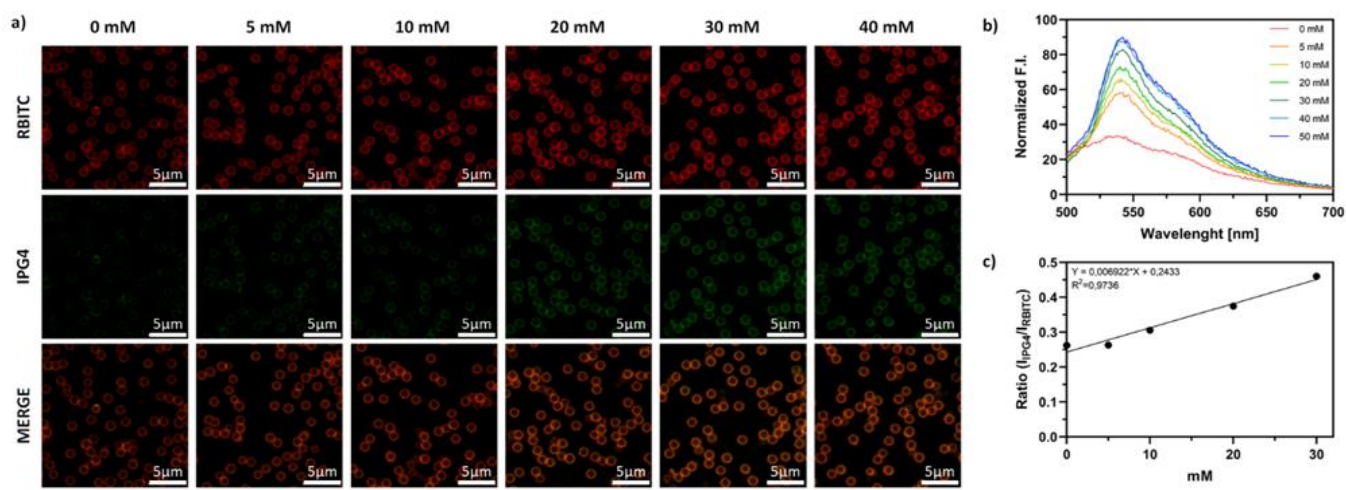

**Figure 4. Calibration of K$^+$-sensors. a)** Representative CLSM micrographs of SiO$_2$@RBITC-IPG4 microparticles incubated in MES buffer (50 mM) with adjusted concentrations of KCl (0, 5, 10, 20, 30, 40 mM). RBITC red channel ($\lambda_{ex}$ 555 nm, $\lambda_{em}$ 600-700 nm), IPG4 green channel ($\lambda_{ex}$ 488 nm, $\lambda_{em}$ 500-600 nm), ZEISS LSM 700 objective 63X, zoom 2, scale bars 5 µm; **b)** emission spectra of SiO$_2$@RBITC-IPG4 microparticles incubated in MES buffer (50 mM) with adjusted concentrations of KCl (0, 5, 10, 20, 30, 40, 50 mM) with $\lambda_{ex}$ 488 nm; **c)** calibration curve of SiO$_2$@RBITC-IPG4 based on CLSM image analysis, on the top the plot with K$^+$ concentration from 0 to 30 mM, standard error shown as SEM bar.

**Physicochemical characterization**

The physicochemical properties of the three synthesized microparticles, namely SiO$_2$@IPG4, SiO$_2$@Cy3-IPG4 and SiO$_2$@RBITC-IPG4, were investigated by means of DLS.

As reported in Table 1, microparticles displayed a hydrodynamic diameter between 1.5 to 2.0 µm, with relatively low polydispersity index. As expected, the three systems showed a slightly negative surface charge, due to the covalent bond of negatively charged fluorophores. Size and morphology of the microsensors were also characterized through SEM analysis (Figure 3). All of the three types of silica microparticles exhibited spherical shape, with a homogeneous smooth surface and a good level of monodispersity. The measured diameter of SiO$_2$@IPG4 MPs, extracted from the SEM image analysis, resulted being 1.374 ± 0.016 µm (Figure 3a-c), which was comparable to the hydrodynamic diameter obtained by DLS analysis. SiO$_2$@Cy3-IPG4 MPs appeared to be the most polydispersed sensors, with a mean diameter of 1.195 ± 0.039 µm (Figure 3d-f). The SiO$_2$@RBITC-IPG4 MPs were the smallest of the three systems, with a mean diameter of 1.171 ± 0.018 µm (Figure 3g-i). Overall, a high level of reproducibility in the microparticles synthesis, with a size in the range of 1.0-1.5 µm, was achieved. The small fluctuations in diameter and size distribution may be related to small differences in the reaction time and the molecular weight of the dyes bond to the MPs surface, that might affect the Stöber condensation, which is known to be highly sensitive to slight variations in reaction condition [32].

**Calibration of the ratiometric fluorescent K$^+$ sensing microparticles by CLSM-imaging**

The fluorescent ratiometric microsensors were calibrated by means of CLSM in MES buffers with a known concentration of K$^+$, between 0 to 30 mM, with the increase in the fluorescence intensity of the green channel. Such lower sensitivity, compared to the fluorimetric calibration, might be related to the different sensitivity of the CLSM, which was confirmed by the LOD calculated for the two method. The LOD obtained with the spectrofluorimetric calibration is 1.7 mM, while the LOD for the CLSM calibration is 9.51 mM. On the contrary, the calibration performed on the SiO$_2$@Cy3-IPG4 microsensors (see Figure S4) was not reproducible as the SiO$_2$@RBITC-IPG4 ($R^2$=0.1953 and $R^2$=0.9736, respectively). In addition, the SiO$_2$@Cy3-IPG4 microparticles, despite being sensitive, were not homogeneously functionalized and fluorescent (see Figure S5), with only few particles displaying a fluorescence signal intense enough to allow the image analysis. Such issue might be related to the lower quantum yield of the Cy3 compared to RBITC and with a different molecular structure, that might influence the microsensors synthesis. The higher molecular weight, steric hindrance and negative charge of Cy3 might negatively affect the Stöber reaction involved in the microparticles synthesis. Since SiO$_2$@RBITC-IPG4 MPs proved to be the best working system, we tested the stability of the microsensors over 7 days. In detail, the ratiometric SiO$_2$@RBITC-IPG4 MPs were incubated with different concentration of K$^+$ and the emission peak of the sensing system was monitored over one week in order to record fluctuations occurring during the storage. The results, reported in Figure S6, showed that the ratiometric signal of the MPs remained stable at each time point collected in the range from 0 to 40 mM of [K$^+$]. Therefore, this evidence highlighted the ability of the SiO$_2$@RBITC-IPG4 MPs to monitor its target analyte and supported the application of the optimized microparticles as valuable and robust sensing tool for K$^+$ detection.

**Biocompatibility evaluation**

The biocompatibility of the ratiometric microsensors was assessed through in vitro cell viability assay. We selected four different cell lines, three derived from different human tumor tissues, in order to assess the toxicity on a broad spectrum of tumor models, and a non-cancerous cell lines: breast cancer cells (MCF-7), human pancreatic adenocarcinoma cells (PANC-1), malignant melanoma cells (SK-MEL2) and murine fibroblasts (3T3). Cells were treated with three increasing concentrations (0.05-0.1-0.3 mg/ml) of $K^+$ sensing microparticles and their cytotoxic effect was evaluated through CellTiter-Glo® Cell Viability Assay. Graphs reported in Figure 5b-d indicate the cell growth measurements derived from the luminescent signal that is proportional to the amount of ATP produced by metabolically active cells after their incubation with CellTiter-Glo® reagent at 0, 24 and 48 hours of treatment. Treated cells showed little to none toxic effect due to the presence of the microsensors. However, MCF-7 and SKMEL-2 cells reported a low toxic effect at highest concentrations, after 24- and 48-hours treatment, but the cell growth percentage never reached values lower than 70%. These results suggest that the $SiO_2$@RBITC-IPG4 microsensors are biocompatible and could be employed in sensing $K^+$ in in vitro cell culture systems.

**Application of microsensors for live-cells $K^+$ detection**

Potassium ions play different roles in biological processes, such as in regulation of cellular electrolyte metabolism, transport of nutrients and nerve transmission [35]. Moreover, dysregulated $K^+$ fluctuations are early signals of diseases, including electrolyte disorders, diabetes and cancer [36]. Therefore, the application of ratiometric optical microsensors in *in vitro* cell culture models could provide great benefits for monitoring cellular metabolism and responses to therapy in pathological conditions, such as cancer.

In this context, the capability of the $SiO_2$@RBITC-IPG4 microsensors to monitor variations of [$K^+$] were tested in a 2D cell culture system of malignant melanoma, SKMEL-2. In order to induce $K^+$ efflux across cell membrane, cells were stimulated with nigericin (1 µM for 1 hour), a $K^+$ ionophore that can form complexes with $K^+$ leading to a decrease in intracellular $K^+$ concentration [37,38]. CLSM acquisitions reported in Figure 5a showed a decrease in the fluorescence intensity of IPG4 dye in the unstimulated cells (CTRL) resulting in an orange overlay with the RBITC reference dye, indicating the physiological concentration of the extracellular $K^+$. On the contrary, in the treated condition, an increase in the fluorescence intensity of IPG4 was observed, indicating the efflux of the intracellular $K^+$ from cells. From the image analyses, the $K^+$ concentration in the CTRL acquisition was determined as 26 ± 0.25 mM. After 1 hour of nigericin treatment, the $K^+$ concentration in the extracellular environment has increased up to 34 ± 0.25 mM. A comparative study based on the fluorometric measurements of the cell culture supernatant collected from nigericin-treated or untreated cells and incubated with $K^+$ sensing microparticles was reported in **Figure S7**. According to the fluorometric analysis, these data showed an increase in the $K^+$ concentration in the cell culture media collected from nigericin-treated cells compared to untreated cells. The $K^+$ concentration in the CTRL was determined as 5.12 ± 0.075 mM; on the contrary, after stimulus with nigericin, the $K^+$ concentration in the culture media has increased up to 7.85 ± 0.06 mM. These results suggested that the $SiO_2$@RBITC-IPG4 microsensors were suitable for sensing $K^+$ with high spatial and temporal resolution in *in vitro* cell culture models.

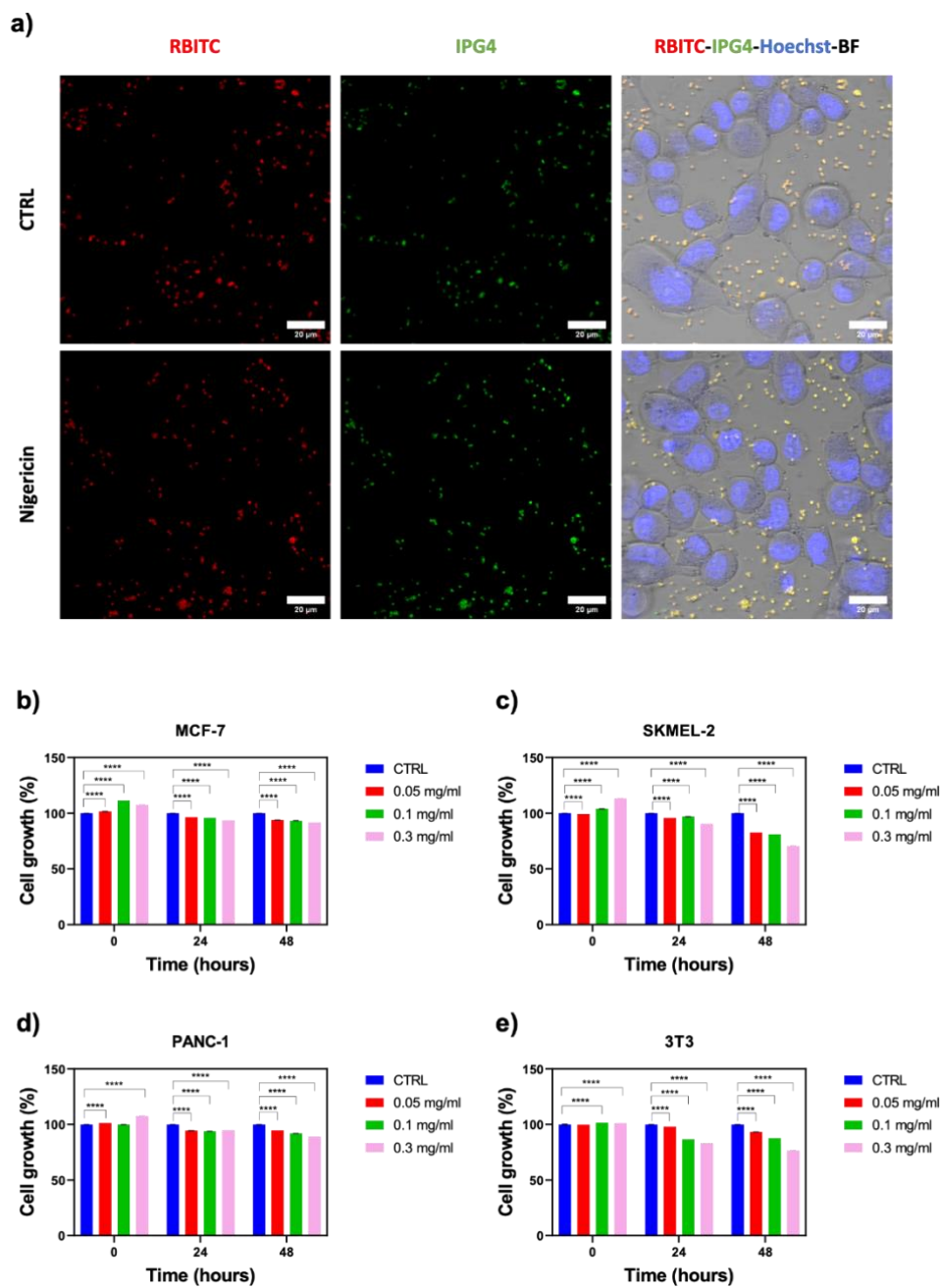

**Figure 5. Application of microsensors in biological systems: a)** CLSM micrographs showing the microsensors properties after an induced pharmacological stimulus. SKMEL-2 cells were seeded in a 4-well chamber slide IBIDI and treated with Nigericin at the concentration of 1 µM for one hour, and the fluorescence emission of the SiO$_2$@RBITC-IPG4 microparticles was recorded by means of CLSM. RBITC red channel ($\lambda_{ex}$ 555 nm, $\lambda_{em}$ 600-700 nm), IPG4 green channel ($\lambda_{ex}$ 488 nm, $\lambda_{em}$ 500-600 nm), Hoechst blue channel ($\lambda_{ex}$ 405 nm, $\lambda_{em}$ 415-500 nm). Images acquisitions were obtained with ZEISS LSM 980 (ZEISS, Germany), Objective 40X Oil-immersion; The microsensors cytotoxicity were tested on **b)** MCF-7, **c)** SKMEL-2, **d)** PANC-1 tumor cells and **e)** 3T3 fibroblasts at three different concentrations (0.05-0.1-0.3 mg/ml) through CellTiter- Glo® Cell Viability Assay. The cell growth percentage was derived from the luminescent signal that is proportional to the amount of ATP produced by metabolically active cells after their incubation with CellTiter-Glo® reagent at 0, 24 and 48 hours of treatment. Values are the means (±SE) of three independent experiments.

## Conclusions

The use of non-invasive tools for inorganic cations monitoring in biologicals samples is a widely explored field and nanotechnology leads the way to new possibilities in term of resolution and accuracy.

In this context our microsensors offer a reliable tool for monitoring K$^+$ variations in the extracellular environment by imaging techniques and spectrofluorimetric analyses. Despite showing sensitivity in the millimolar range of K$^+$ concentration, this is more than appropriate to perform analyses in biological systems.

We designed, synthesized and characterized a ratiometric fluorescent microsensor that ~~is~~ was suitable for the mapping of K$^+$ by means of high-resolution imaging techniques, such as CLSM. We cho~~o~~se silica as the starting material for our microsensor due to its inertness, biocompatibility, flexibility in the synthesis and tailorability of the physical properties. We selected the IPG4 as the K$^+$ sensitive probe, because of its sensitivity and selectivity. We accomplished to conjugate IPG4 on organically modified silica by amide bond formation without affecting its fluorescence and sensitivity. Notably, the binding of a reference dye, required to produce ratiometric K$^+$ sensors, resulted to be critical in preserving the sensitivity of IPG4 versus K$^+$ ions. The best result was obtained by using IPG4 coupled with rhodamine B isothiocyanate as the reference dye. The resulting fluorescent ratiometric microsensors displayed sensitivity in the range between 0 to 40 mM, which is optimal for extracellular tracking, considering that the interstitial K$^+$ concentration is always kept in this range in physiological conditions. Then, the biocompatibility of the K$^+$ sensors and their sensing properties were tested and confirmed in in vitro tumor cell culture models resulting suitable for use in extracellular K$^+$ detection. The developed sensing platform could be used to investigate the role of K$^+$ dysregulation in in vitro models of human diseases. A better understanding of how K$^+$ extracellular fluxes affect physiological process might indeed help the study of pathological conditions such as cancer, epilepsy and neuromuscular disorders. One challenge for *in vivo* imaging with the presented sensor system is the emission of fluorescence in the visible light spectrum, which suffers from limited tissue penetration and potential interference from the autofluorescence of living cells. Therefore, while systemic administration in whole-animal or whole-organ tracking may not be suitable, this ratiometric K$^+$ sensor holds promise for applications in skin monitoring. Potassium dysregulation is a critical factor in skin responses to injury, particularly in wound healing processes. A notable example is skin burns, which are often associated with hyperkalemia. Given the well-documented biocompatibility of silica microparticles and with further optimization of the sensor formulation, these microparticles could become a valuable tool for tracking potassium dynamics in the skin.

## Conflicts of interest

The authors declare no competing interests.

## Data Availability Statement

The data supporting this article have been included as part of the Supplementary Information.

## Acknowledgements

This work was supported by the European Research Council (ERC) under the European Union's Horizon 2020 research and innovation program ERC Starting Grant "INTERCELLMED" (contract number 759959), the European Union's Horizon 2020 research and innovation programme under grant agreement No. 953121 (FLAMIN-GO), the Associazione Italiana per la Ricerca contro il Cancro (AIRC) (MFAG-2019, contract number 22902), the "Tecnopolo per la medicina di precisione" (TecnoMed Puglia) - Regione Puglia: DGR n.2117 of 21/11/2018, CUP: B84I18000540002), the Italian Ministry of Research (MUR) in the framework of the National Recovery and Resilience Plan (NRRP), "NFFA-DI" Grant (B53C22004310006), "I-PHOQS" Grant (B53C22001750006) and under the complementary actions to the NRRP, "Fit4MedRob" Grant (PNC0000007, B53C22006960001) funded by NextGenerationEU and the PRIN 2022 (2022CRFNCP_PE11_PRIN2022) funded by European Union – Next Generation EU.